\documentclass[11pt]{article}
\usepackage{amssymb}
\usepackage{cite}
\usepackage{mathtools}
\usepackage{graphicx}
\usepackage{caption}
\usepackage{subcaption}
\usepackage{amsmath}
\usepackage{amsthm}

\begin{document}
\thispagestyle{empty}
\begin{center}
{\Large\bf Testing biodiversity using inhomogeneous summary statistics}\\[.4in]

\noindent
{\large M.C.~de~Jongh and M.N.M.~van~Lieshout}\\[.1in]
\noindent
{\em CWI, P.O.~Box 94079, NL-1090 GB  Amsterdam\\
Department of Applied Mathematics, University of Twente, \\
P.O.~Box 217, NL-7500 AE Enschede\\
The Netherlands}\\[.1in]
\end{center}

\begin{verse}
{\footnotesize
\noindent
{\bf Abstract}\\
\noindent
McGill's theory of biodiversity is based upon three axioms: individuals of the same species cluster together, many rare 
species co-exist with a few common ones and individuals of different species grow independently of each other. Over the 
past decade, classical point pattern analyses have been employed to verify these axioms based on the false assumption of
stationarity. In this paper, we use inhomogeneous versions of the classical summary statistics for spatial point patterns 
to assess the validity of McGill's first and third axioms for data obtained from a 50 hectare plot on Barro Colorado 
Island. \\[0.1in]

\noindent
{\em Keywords \& Phrases:}
biodiversity, clustering within species, independence 
between species, inhomogeneous $J$-function, inhomogeneous $K$-function, 
spatial point pattern. \\[0.1in]

}
\end{verse}

\bibliographystyle{plain}

\section{Introduction}

The set of laws that govern the distribution and co-existence of species is among the most popular topics of research 
in the field of ecology \cite{Ricklefs}. In 1934, biologist G.F.\ Gause stated the \textit{competitive exclusion 
principle\/}, which asserts that two species cannot stably co-exist if they compete for the same means of living 
\cite{Gause}. Yet in tropical rain forests or coral reefs a vast number of different species are encountered within 
small areas \cite{Gentry, Huisman, Losos}. Over the years, various attempts have been made to explain such observed 
patterns of biodiversity \cite{Chesson, Clements, Gleason, Hubbell1, Tilman, Wright}. In 2010, biologist McGill set 
out to unite six different theories into one \cite{McGill}. He established three axioms that would constitute the 
basis of his unified theory of biodiversity: individuals of the same species cluster together, many rare species 
co-exist with a few common ones and individuals of different species grow independently of each other. 

McGill's axioms have been subjected to careful scrutiny over the past decade. Various studies showed that interactions 
within a species are generally much stronger than inter-specific interactions \cite{Lieberman, Perry,  Wiegand2}. In 2012, 
Wiegand {\em et al} used summary statistics for spatial point processes to examine the validity of the third of McGill's 
axioms \cite{Wiegand3}. The results of applying these statistics to data from three tropical rain forests reveal that 
the strength of inter-specific interactions fades with species richness. A stumbling block in the approach, though, is 
the fact that the summary statistics used assume stationarity of the process, which in this context is not likely to 
hold. Indeed, the authors observe that classic point process analyses of bivariate spatial patterns are challenging 
'because they require complete mapping and because of difficulties in teasing apart two major, yet contrasting factors: 
habitat association and direct species interactions.' The first of these difficulties was addressed in \cite{VanLieshout1},
where summary statistics for random measures were used to eliminate the requirement of a complete mapping. Here, we 
address the second problem using recent advances in spatial statistics on the development of inhomogeneous analogues 
of the classical summary statistics \cite{Baddeley1, VanLieshout2}. More specifically, we will use appropriate 
inhomogeneous summary statistics to investigate the validity of McGill's first and third axiom for data on the spatial 
distribution of stems of a large number of woody trees and shrub species measuring at least 1cm in diameter found in a 
$50$ hectare plot on Barro Colorado Island, Panama \cite{Hubbell2}. This plot was established in 1980 by researchers 
from the Smithsonian Tropical Research Institute and Princeton University and has been analysed in numerous studies 
\cite{Volkov, Wiegand3, Waagepetersen}. The first census was conducted in 1982 after which new censuses were carried 
out every five years since 1985. Here, we will use data from the $8$-th 
census \footnote{{\tt https://doi.org/10.15146/5xcp-0d46}}.

The plan of this paper is as follows. In Section~\ref{S:sumstats}, we recall the definitions of the inhomogeneous 
$K$- and $J$-function and their cross versions. Section~\ref{S:axiom1} is devoted to intra-specific interactions, 
whilst Section~\ref{S:axiom3} deals with inter-specific interactions. We will summarize our findings in 
Section~\ref{S:conclusions}.

\section{Summary statistics}
\label{S:sumstats}

Typically, spatial data such as those gathered on Barro Colorado Island are modelled as a so-called point process 
\cite{Diggle1}, a random mechanism for generating finite patterns in some bounded observation window $W \subset 
\mathbb{R}^d$. Let $X$ denote such a point process in $\mathbb{R}^d$. Then $X$ is called \textit{stationary} if 
its probability distribution is invariant under translations. Intuitively speaking, this means that the output of 
the random mechanism looks the same in terms of the point density and the inter-point interactions from every viewpoint 
in $\mathbb{R}^d$. 

Structural information about point processes is often expressed by means of summary statistics, as described in 
detail by Diggle \cite{Diggle1}. The well-known $K$-function for example is proportional to the expected number of 
further points seen within a given distance of a typical point. For radius $ r \geq 0$, it can be estimated by
\[
\hat{K}(r) = \dfrac{1}{|W|} \sum\limits_{x_i \in X \cap W} 
\sum\limits_{x_j \in X \cap W \setminus \{x_i\}} \dfrac{w_{x_i,x_j} 
  1 (\parallel x_i - x_j \parallel \leq r)}{ (\hat \lambda)^2}, 
\]
where $\hat \lambda$ is the average number of points per unit area. Here $w_{x_i, x_j}$ is a weight factor to 
compensate for edge effects \cite{Baddeley4,Diggle1} and $1 (\parallel x_i - x_j \parallel \leq r)$ takes the value 
one if the distance between $x_i$ and $x_j$ is at most $r$, zero otherwise. If $X$ is not stationary, the abundance 
of points varies over space. Thus, to estimate the \textit{inhomogeneous $K$-function} proposed by Baddeley {\em et al} 
\cite{Baddeley1}, $\hat \lambda$ is replaced by an estimate $\hat \lambda(x)$ of the spatially varying intensity function. 
Doing so, we obtain 
\begin{equation} \label{e:K-inhom}
\hat{K}_{\text{inhom}}(r) = 
\dfrac{1}{|W|} \sum\limits_{x_i \in X \cap W} \sum\limits_{x_j \in X \cap W \setminus \{x_i\}} 
\dfrac{w_{x_i,x_j} 1 (\parallel x_i - x_j \parallel \leq r)}
{\hat \lambda(x_i) \hat\lambda(x_j)}.
\end{equation}
The term $\hat \lambda(x_i) \hat \lambda(x_j)$ accounts for the habitat associations, so that $K_{\rm{inhom}}$ is 
solely a measure of the direct interactions within a species.

Inhomogeneous extensions of the empty space function $F(r)$ and the nearest-neighbour distance distribution function 
$G(r)$  were constructed in a similar manner by Van Lieshout \cite{VanLieshout2}. The inhomogeneous $J$-function 
$J_{\rm{inhom}}$ is simply the ratio of $1-G_{\rm{inhom}}$ and $1 - F_{\rm{inhom}}$, provided the latter is non-zero.

In studying inter-species interactions, we make use of the cross versions of summary statistics that describe 
associations between points of different types. The cross $K$-function, $K_{1,2}(r)$, for example, is proportional to 
the number of points of type $2$ seen within distance $r$ from a typical point of type $1$ and can be estimated in a 
similar fashion as the univariate $K$-function upon weighting a point of type $l \in \{ 1, 2 \}$ by $\hat \lambda_l$, 
the average number of points of type $l$ per unit area.  To estimate the inhomogeneous cross $K$-function, simply allow 
$\hat \lambda_l(x)$ to be spatially varying \cite{MollerWaagepetersen}. An inhomogeneous cross version of the $J$-function 
was introduced in \cite{VanLieshout2} and studied in \cite{Cronie1}. For further technical details and underlying 
assumptions, we refer to the literature cited above.

\section{Clustering within species}
\label{S:axiom1}

\begin{figure}[hbt]
\centering
\begin{subfigure}{.45\textwidth}
  \centering
  \includegraphics[width=0.8\linewidth]{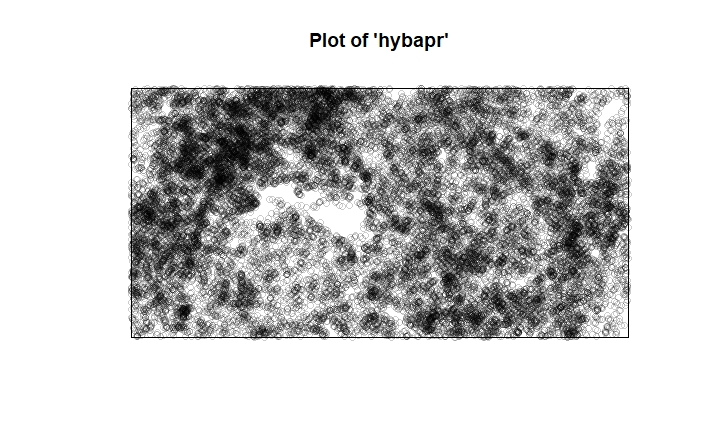}
  \caption{Locations of {\em Hybanthus prunifolius\/} (`hybapr') trees.}
  \label{hybapr}
\end{subfigure}\hspace{0.02\textwidth}
\begin{subfigure}{.45\textwidth}
  \centering
  \includegraphics[width=0.8\linewidth]{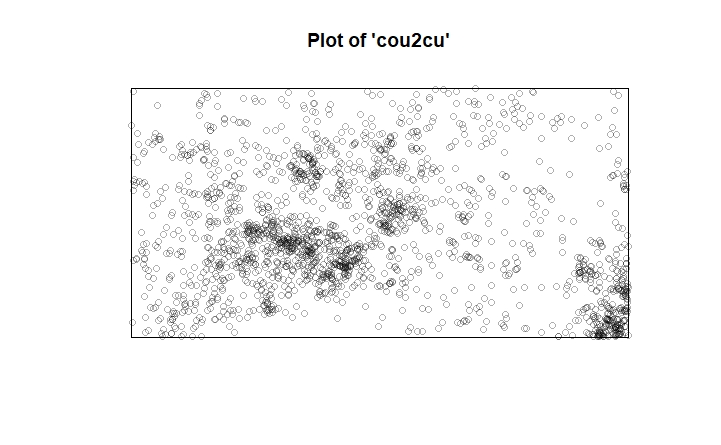}
  \caption{Locations of {\em Coussarea curvigemmia\/} (`cou2cu') trees.}
  \label{cou2cu}
\end{subfigure}\hspace{0.02\textwidth}
\begin{subfigure}{.45\textwidth}
  \centering
  \includegraphics[width=0.8\linewidth]{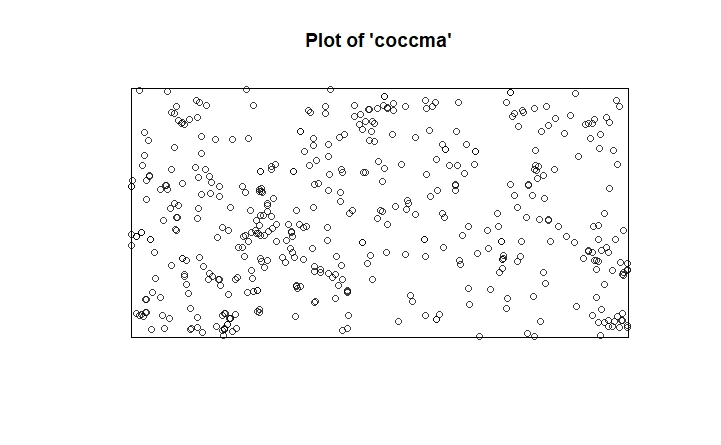}
  \caption{Locations of {\em Coccoloba manzanillensis\/} (`coccma') trees.}
  \label{coccma}
\end{subfigure}\hspace{0.02\textwidth}
\begin{subfigure}{.45\textwidth}
  \centering
  \includegraphics[width=0.8\linewidth]{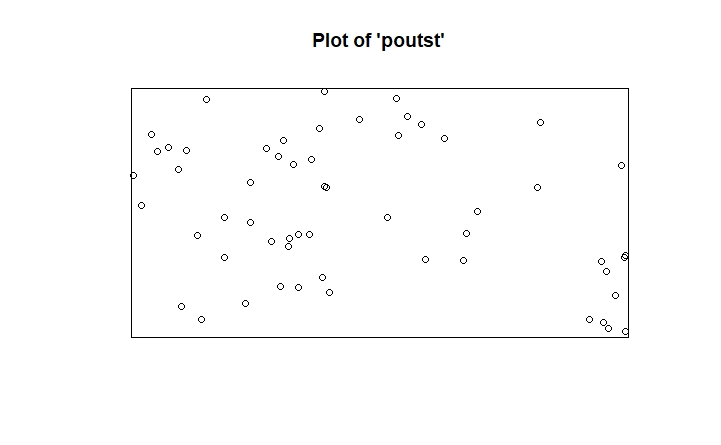}
  \caption{Locations of {\em Pouteria stipitata\/} (`poutst') trees.}
  \label{poutst}
\end{subfigure}
\caption{Locations of alive trees in the eighth census on Barro Colorado Island for some selected species.}
\label{plots}
\end{figure}

In this section, we investigate the validity of McGill's first axiom, which states that individuals from the same 
species tend to cluster together. For illustration purposes, we pick four species ranging from quite rare to most common. 
The locations of the trees of these species are shown in Figure~\ref{plots}.

It is evident from the plots in Figure~\ref{plots} that trees are not scattered in a homogeneous fashion over the study 
region. Therefore, in order to check whether trees of the selected species tend to cluster together, we subject them to 
the inhomogeneous versions of the summary statistics discussed in Section~\ref{S:sumstats}. Since the intensity functions 
of the point processes are unknown, we estimate them using a non-parametric Gaussian kernel estimator \cite{Diggle2} with 
local edge correction \cite{VanLieshout3}. The quality of such estimates relies heavily on the choice of the smoothing 
bandwidth. Here we use Cronie and Van Lieshout's criterion \cite{Cronie2}.

\begin{figure}
\centering
\begin{subfigure}{.4\textwidth}
  \centering
  \includegraphics[width=0.8\linewidth]{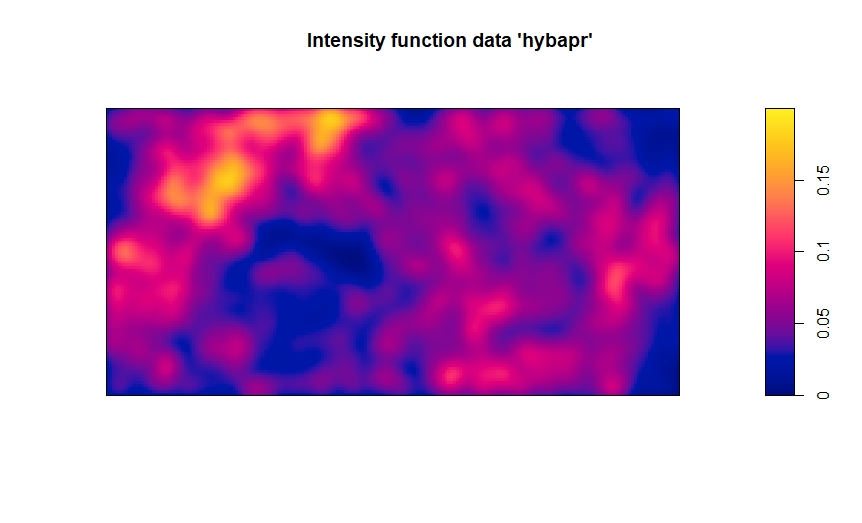}
  \caption{Kernel smoothed intensity function of `hybapr'.}
  \label{lambda_hybapr}
\end{subfigure}\hspace{0.02\textwidth}
\begin{subfigure}{.4\textwidth}
  \centering
  \includegraphics[width=0.8\linewidth]{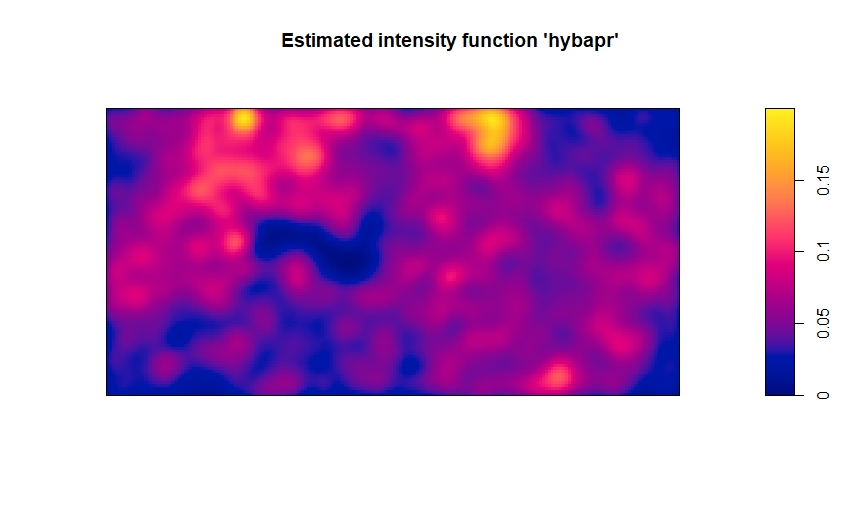}
  \caption{Estimated intensity function for `hybapr'.}
  \label{lambda_hybapr_sim}
\end{subfigure}\\
\begin{subfigure}{.4\textwidth}
  \centering
  \includegraphics[width=0.8\linewidth]{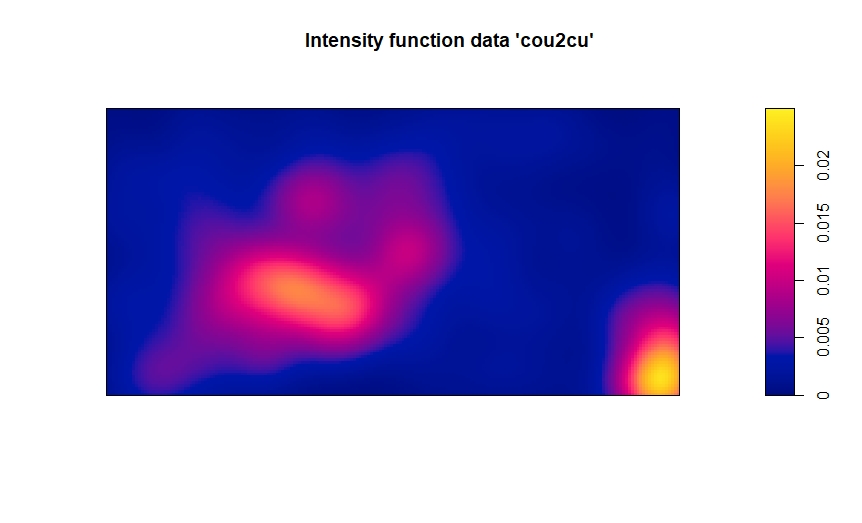}
  \caption{Kernel smoothed intensity function of `cou2cu'.}
  \label{lambda_cou2cu}
\end{subfigure}\hspace{0.02\textwidth}
\begin{subfigure}{.4\textwidth}
  \centering
  \includegraphics[width=0.8\linewidth]{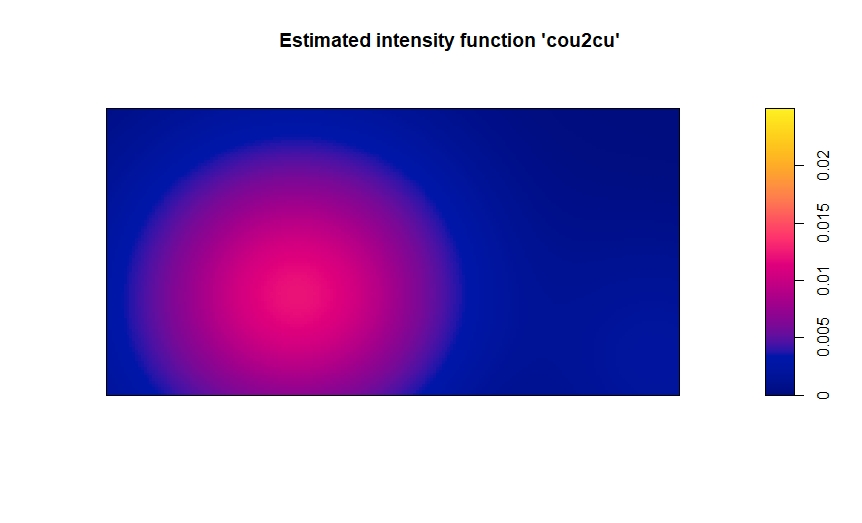}
  \caption{Estimated intensity function for `cou2cu'.}
  \label{lambda_cou2u_sim}
\end{subfigure}\\
\caption{Comparison of estimated intensity functions.}
\label{intfunctions}
\end{figure}

For each of the species, we compare the estimates of the inhomogeneous summary statistics to the ones that correspond 
to an inhomogeneous Poisson process. As in the homogeneous case, $K_{\text{inhom}}^{\text{poiss}}(r) = \pi r^2$. An 
estimated value that is greater than $\pi r^2$ suggests clustering. For the $J$-function, 
$J_{\text{inhom}}^{\text{poiss}}(r) = 1$; values less than one indicate clustering. In order to assess if the deviations 
of the estimated statistics from those of a Poisson process are significant, a range of Monte Carlo tests 
\cite{Baddeley4, Myllymaki} can be conducted. We use the one-sided versions that test against clustered alternatives.

Specifying the intensity function of the null process proves to be difficult. We could follow \cite{Baddeley4} and use 
the observed point pattern to obtain an estimate $\hat{\lambda}(x)$. However, that would violate the principle of not 
using the same data for estimation as for validation. Moreover, under the null hypothesis the simulated patterns are 
realizations of an inhomogeneous Poisson process with intensity function $\hat{\lambda}(x)$, whereas the observed 
pattern is an inhomogeneous Poisson process with unknown intensity function $\lambda(x)$. In order to circumvent this 
issue, we estimate the intensity function using data from the first census which has the least overlap with the observed 
pattern. To reduce the dependence of the estimates on the observed data even further, we remove all trees that 
were still alive at the time of the latest census. The intensity estimate based on the trimmed data is then scaled in 
such a way that the total number of points it predicts in the plot is equal to the actual number of trees counted during 
the latest census. Note that this method is not perfect in that it still uses the observed tree counts. To make sure that 
it is quite a reasonable option, we compare the intensity function thus obtained with a kernel estimate based on the data. 
Both functions are depicted in Figure~\ref{intfunctions} for the species `hybapr' and `cou2cu'. The intensity functions 
of the remaining species can be found in the appendix.

\begin{figure}
\centering
\begin{subfigure}{.45\textwidth}
  \centering
  \includegraphics[width=0.8\linewidth]{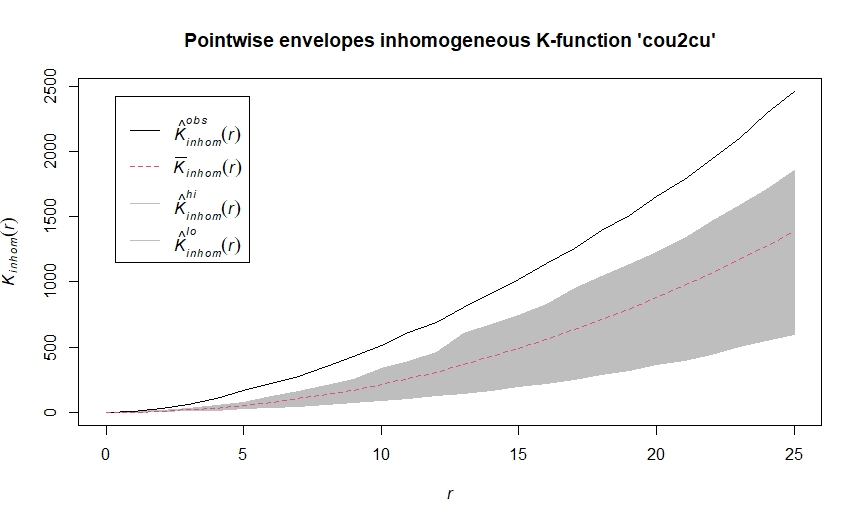}
  \caption{Pointwise envelopes of estimated inhomogeneous $K$-function.}
  \label{Kenvelopes_cou2cu}
\end{subfigure}\hspace{0.02\textwidth}
\begin{subfigure}{.45\textwidth}
  \centering
  \includegraphics[width=0.8\linewidth]{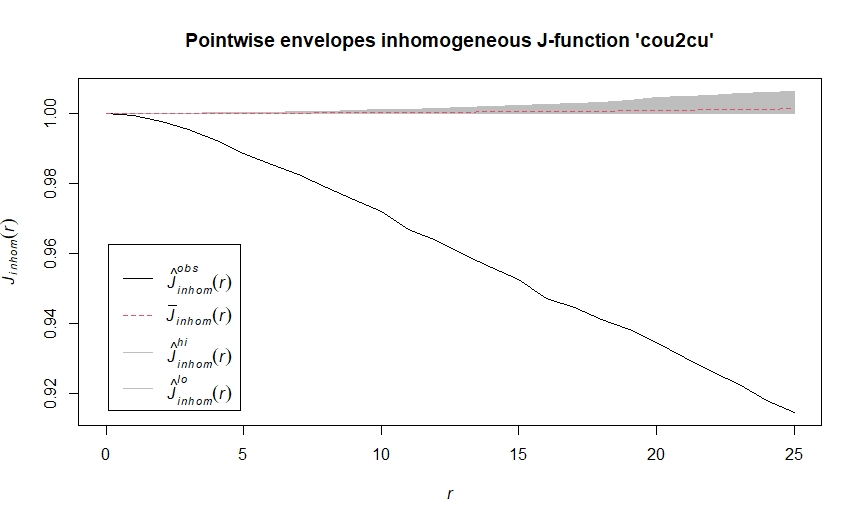}
  \caption{Pointwise envelopes of estimated inhomogeneous $J$-function.}
  \label{Jenvelopes_cou2cu}
\end{subfigure}\hspace{0.02\textwidth}
\caption{Simulation envelopes for estimated inhomogeneous $K$- and $J$-functions for `cou2cu'
based on $19$ simulations.}
\label{envelopes_cou2cu}
\end{figure}

Figure \ref{intfunctions} suggests that the estimates are fairly good. The areas that are rich or barren in vegetation
seem to be roughly in harmony. However, the figures do show some differences that may give rise to distortions in the
test results. For example, the data of the latest census seem to feature a cluster of `cou2cu' trees in the bottom right
corner of the plot which is not captured by the estimated intensity functions.

Pointwise envelopes for the species `cou2cu' are shown in Figure~\ref{envelopes_cou2cu} based on $19$ simulated patterns.
Both summary statistics indicate that `cou2cu' trees cluster together. After all, the estimated $K$-function lies above
the upper envelope and the estimated $J$-function lies beneath the lower envelope over the entire range of $r$. Similar
figures for the remaining species are provided in the appendix.

\begin{table}[hbt]
\begin{center}
 \begin{tabular}{||c||c||c||c||c||} 
 \hline
 Summary & MAD test & DCLF test & Studentized & Directional quantile \\
  statistic  & & & MAD test & MAD test \\
 \hline
 \hline
 K & 0.01 & 0.01 & 0.02 & 0.01\\
 J & 0.01 & 0.01 & 0.01 & 0.01\\
 \hline  
\end{tabular}    
\caption{$P$-values of the MAD test, the DCLF test, the studentized MAD test and the directional quantile MAD test for `cou2cu'.}
\label{testresults1}
\end{center}
\end{table}

To assess whether the clustering is statistically significant, Table \ref{testresults1} lists the $p$-values returned 
by the MAD test, the DCLF test, the studentized MAD test and the directional quantile MAD test \cite{Baddeley4, Myllymaki} 
over the range $[0,25]$. In conducting these tests, we used the envelopes based on $99$ simulations. Since one-sided 
versions of the studentized and the directional quantile MAD test are not available, we use the two-sided versions of 
these tests. The results strongly support clustering of `cou2cu' trees. Tables for the other three species are provided 
in the appendix. The results for `hybapr' and `coccma' are similar to those for `cou2cu'. For `poutst' on the other hand, 
we do not find much evidence for clustering. 

\begin{figure}
\centering
\begin{subfigure}{.45\textwidth}
  \centering
  \includegraphics[width=0.8\linewidth]{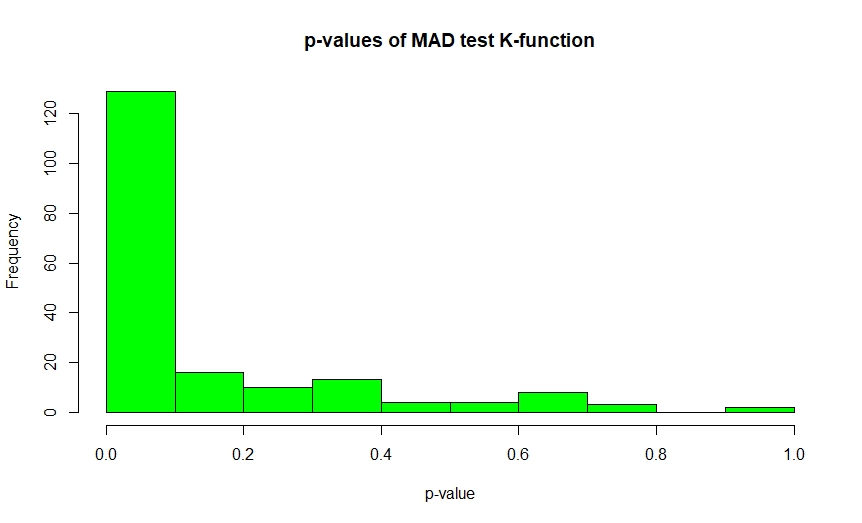}
  \caption{Distribution of $p$-values for the MAD test based on the inhomogeneous $K$-function.}
  \label{pMADKbig_def}
\end{subfigure}\hspace{0.02\textwidth}
\begin{subfigure}{.45\textwidth}
  \centering
  \includegraphics[width=0.8\linewidth]{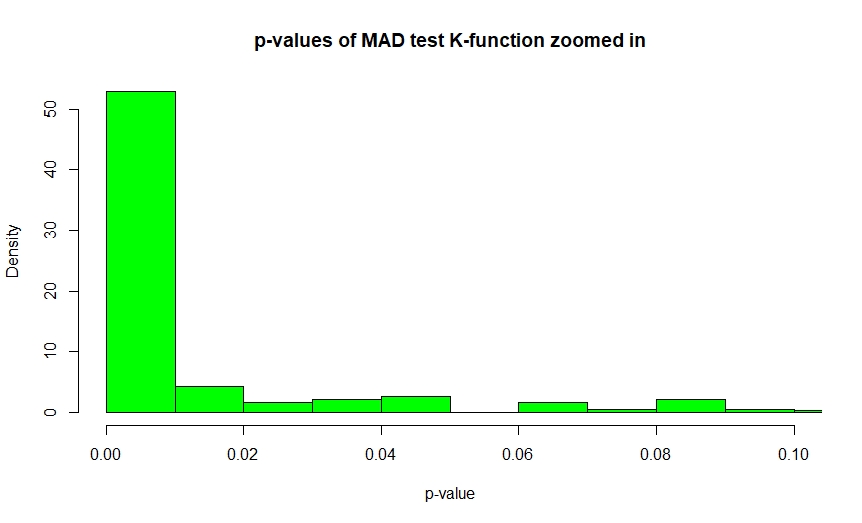}
  \caption{Distribution of zoomed in $p$-values for the MAD test based on the inhomogeneous $K$-function.} 
  \label{pMADKsmall_def}
\end{subfigure}\hspace{0.02\textwidth}
\begin{subfigure}{.45\textwidth}
  \centering
  \includegraphics[width=0.8\linewidth]{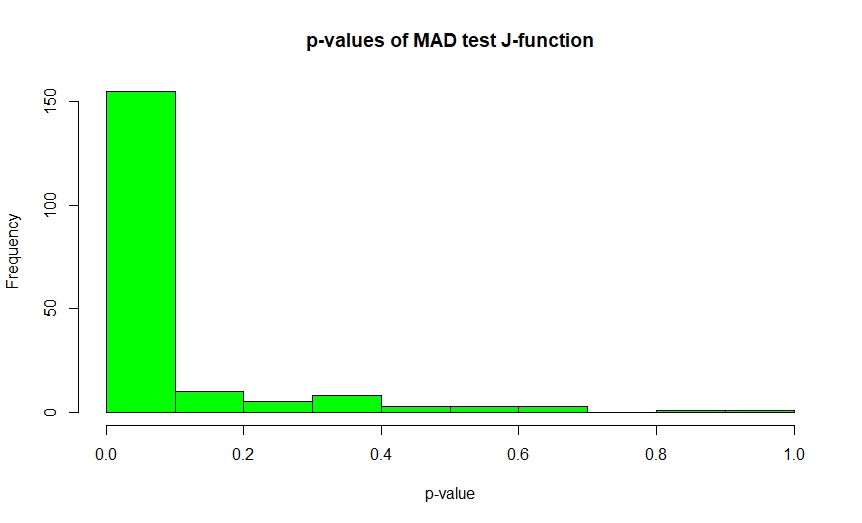}
  \caption{Distribution of $p$-values for the MAD test based on the inhomogeneous $J$-function.}
  \label{pMADJbig_def}
\end{subfigure}\hspace{0.02\textwidth}
\begin{subfigure}{.45\textwidth}
  \centering
  \includegraphics[width=0.8\linewidth]{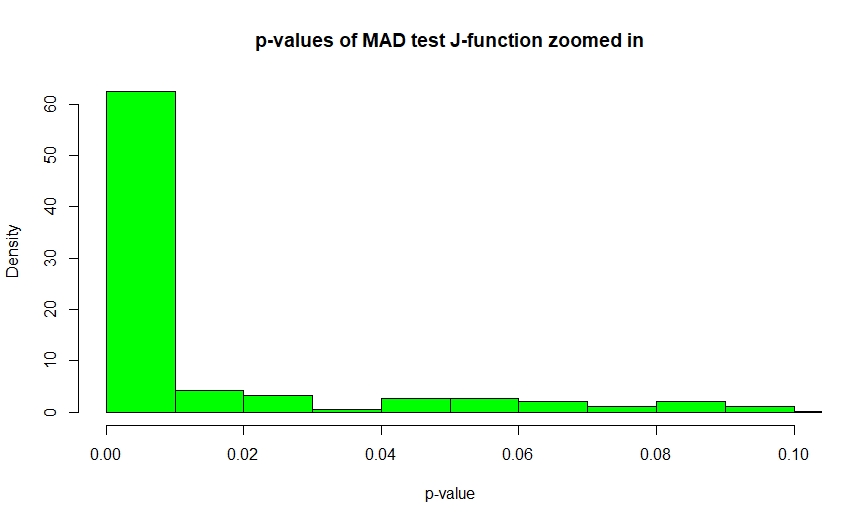}
  \caption{Distribution of zoomed in $p$-values for the MAD test based on the inhomogeneous $J$-function.} 
  \label{pMADJsmall_def}
\end{subfigure}\hspace{0.02\textwidth}
\caption{Distribution of $p$-values for the MAD test.}
\label{generalization1}
\end{figure}

To examine if the selected species are representative for the vegetation in the plot, we carry out the MAD test for each 
species featuring more than $50$ trees. Due to the high number of species, here we use the kernel smoothed data to generate 
samples. Figure~\ref{generalization1} shows the distributions of the obtained $p$-values when using the inhomogeneous 
$K$- and $J$-functions. Both provide convincing evidence for McGill's first axiom: the histograms show distinct peaks 
between $0 $ and $0.01$.

\section{Independence between species}
\label{S:axiom3}

The third of McGill's axioms states that the patterns of trees of different species are independent of each other. 
In order to assess the validity of this axiom, we conduct a Lotwick--Silverman test for independence \cite{Lotwick} 
applied to the inhomogeneous cross $K$- and $J$-functions. Recall that when species $i$ and $j$ are independent, 
$K_{\text{inhom; i,j}}(r) = \pi r^2$; larger values suggest positive association, smaller values negative association 
between the species. Similarly,  $J_{\text{inhom; i,j}}(r) = 1$ when the species are independent. Smaller values suggest 
that trees of species $j$ tend to cluster around those of species $i$, larger ones that trees of type $j$ tend to avoid 
those of type $i$. The idea behind the test is that under the null hypothesis of independence, shifting the pattern and 
the intensity function of one of the species with respect to those of the other leaves the marginal structures intact 
and affects only the inter-species interactions. Since such a shift might move some points out of the plot, a torus 
correction is applied.  

\begin{figure}
\centering
\begin{subfigure}{.45\textwidth}
  \centering
  \includegraphics[width=0.8\linewidth]{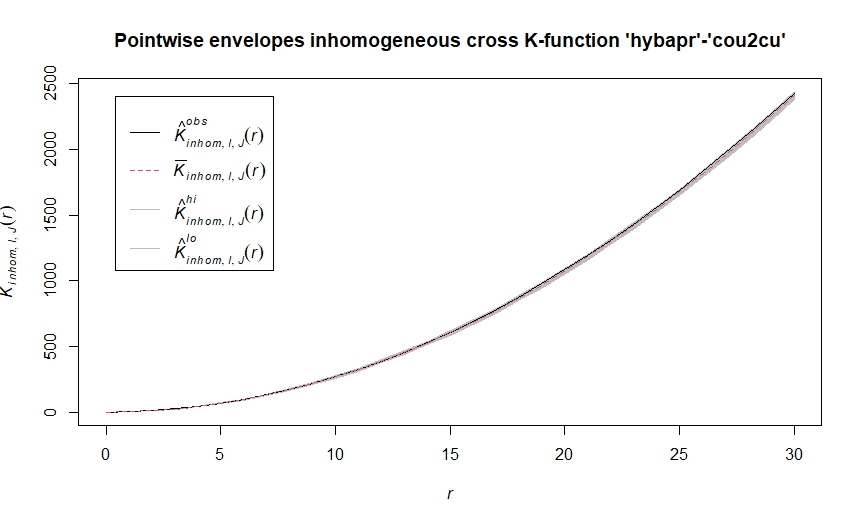}
  \caption{Pointwise envelopes of estimated inhomogeneous cross $K$-function.}
  \label{Kcrossenvelopes}
\end{subfigure}\hspace{0.02\textwidth}
\begin{subfigure}{.45\textwidth}
  \centering
  \includegraphics[width=0.8\linewidth]{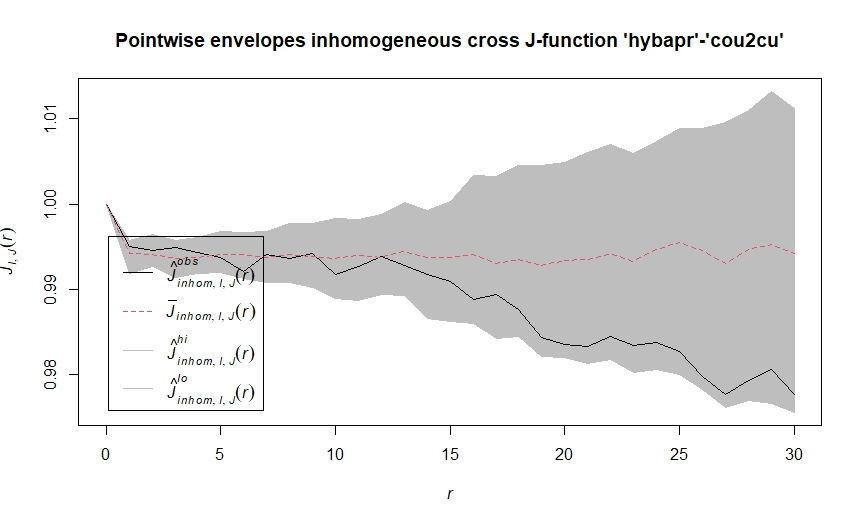}
  \caption{Pointwise envelopes of estimated inhomogeneous cross $J$-function.}
  \label{Jcrossenvelopes}
\end{subfigure}\\
\caption{Simulation envelopes for estimated inhomogeneous cross $K$- and $J$-functions from `hybapr' to `cou2cu'
based on $99$ simulations.}
\label{crossenvelopes}
\end{figure}

Figure \ref{crossenvelopes} shows the pointwise envelopes of the estimated inhomogeneous cross $K$- and $J$-functions 
for the species pair `hybapr'-`cou2cu', each based on $99$ torus translations of the first species with respect to the 
other. The graphs do not give us any reason to doubt the hypothesis of independence, as the estimated summary statistics 
for the data lie mostly within the envelopes. The appendix contains pointwise envelopes for various other species pairs. 
Note that for some species pairs, we find hints of a positive association.  

Table \ref{testresults2} shows the $p$-values returned by the two-sided MAD test, the DCLF test, the studentized MAD 
test and the directional quantile MAD test applied to the inhomogeneous cross $K$- and $J$-functions for the pair 
`hybapr' -- `cou2cu' over the range $[0,30]$. In conducting these tests, we again used the envelopes based on $99$ 
simulations.  All $p$-values are $5$ percent or higher and there is no reason to reject the independence hypothesis.
In the appendix, similar tables are given for several other species pairs.

\begin{table}[hbt]
\begin{center}
 \begin{tabular}{||c||c||c||c||c||} 
 \hline
 Summary statistic & MAD test & DCLF test & Studentized & Directional quantile \\
  & & & MAD test & MAD test \\
 \hline
 \hline
 K & 0.28 & 0.30 & 0.60 & 0.53\\
 J & 0.06 & 0.06 & 0.20 & 0.08\\
 \hline  
 \end{tabular}    
\caption{$P$-values of the MAD test, the DCLF test, the studentized MAD test and the directional quantile MAD test for `hybapr' -- `cou2cu'.}
\label{testresults2}
\end{center}
\end{table} 

\begin{figure}
\centering
\begin{subfigure}{.45\textwidth}
  \centering
  \includegraphics[width=0.8\linewidth]{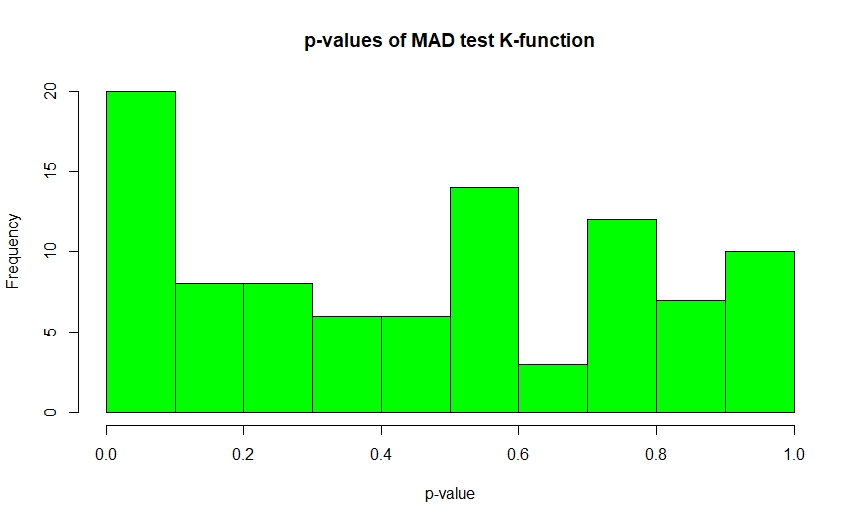}
  \caption{Distribution of $p$-values for MAD test based on the inhomogeneous cross $K$-function.}
  \label{Generalization3Kbig}
\end{subfigure}\hspace{0.02\textwidth}
\begin{subfigure}{.45\textwidth}
  \centering
  \includegraphics[width=0.8\linewidth]{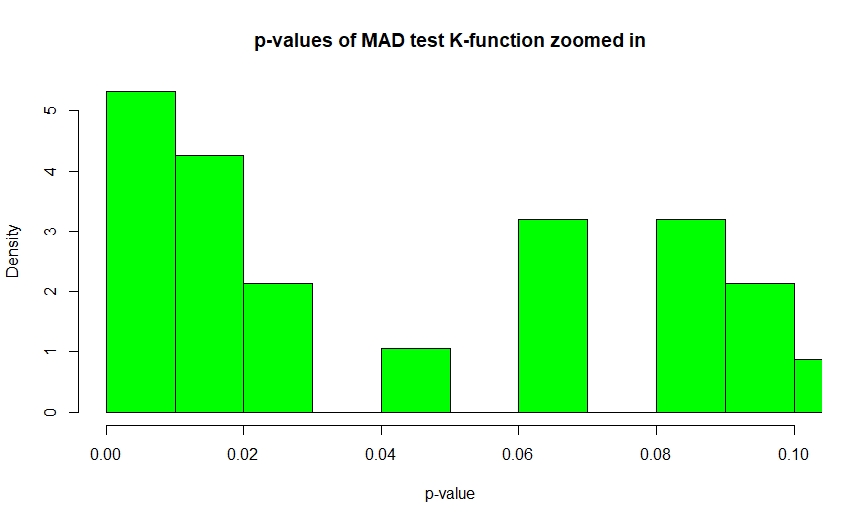}
  \caption{Distribution of zoomed in $p$-values for MAD test based on the inhomogeneous cross $K$-function.}
  \label{Generalization3Ksmall}
\end{subfigure}\hspace{0.02\textwidth}
\begin{subfigure}{.45\textwidth}
  \centering
  \includegraphics[width=0.8\linewidth]{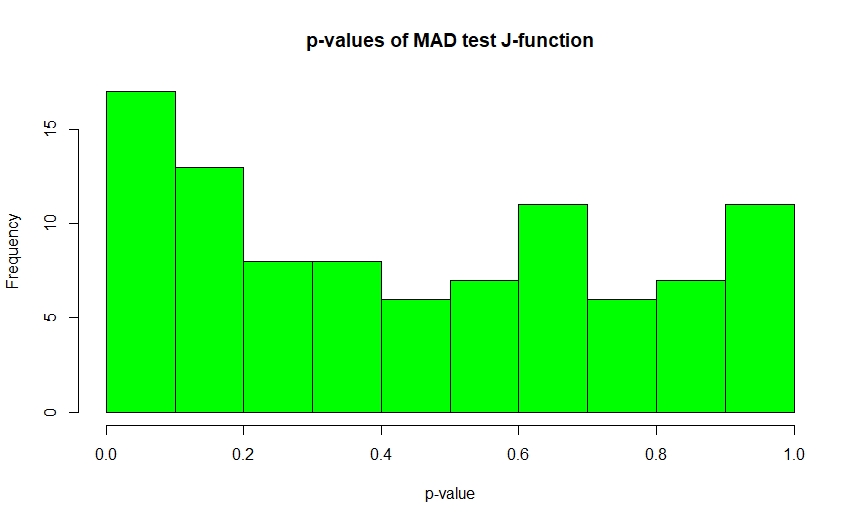}
  \caption{Distribution of $p$-values for MAD test based on the inhomogeneous cross $J$-function.}
  \label{Generalization3Jbig}
\end{subfigure}\hspace{0.02\textwidth}
\begin{subfigure}{.45\textwidth}
  \centering
  \includegraphics[width=0.8\linewidth]{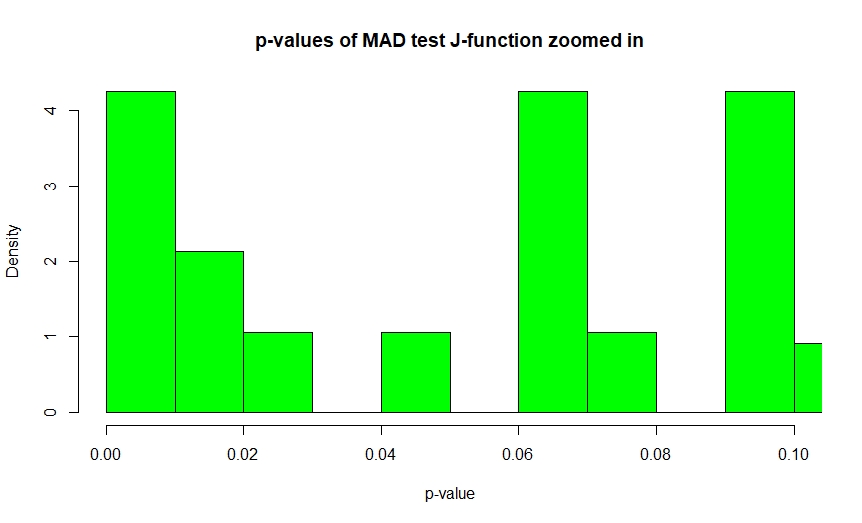}
  \caption{Distribution of zoomed in $p$-values for MAD test based on the inhomogeneous cross $J$-function.}
  \label{Generalization3Jsmall}
\end{subfigure}\hspace{0.02\textwidth}
\caption{Distribution of $p$-values for the MAD test.}
\label{generalization2}
\end{figure}

In order to gain more insight in the validity of the third axiom, we examine the total pool of species pairs. We 
restrict ourselves to patterns containing more than $50$ trees and pair them randomly. We then estimate the 
inhomogeneous cross $K$- and $J$-functions and apply the MAD test as before. The distributions of the obtained 
$p$-values are shown in Figure~\ref{generalization2}. Both for the $K$- and for the $J$-function, over $85$ percent 
of the $p$-values exceed 0.05. Hence, the third axiom seems to hold for the majority of pairs.

\section{Conclusions}
\label{S:conclusions}

In this paper, we confirmed the validity of the first and third of McGill's axioms of biodiversity using data on 
the spatial distribution of trees found in a $50$ hectare plot on Barro Colorado Island \cite{Hubbell2} by means 
of summary statistics that do not rely on the assumption of stationarity. 

Although the results of this study are highly indicative of the correctness of McGill's axioms, we should be heedful 
not to draw too presumptuous conclusions. More and different kinds of ecosystems should be investigated. Furthermore, 
recall that the results of the Monte Carlo tests depend on various choices, such as the type of edge correction used, 
the estimated intensity function, the choice of bandwidth and the range of interaction considered. The robustness of 
the results with respect to these choices should also be investigated.

\mbox{}

\section*{Acknowledgements}

Thanks are due to Professor Richard J.\ Boucherie for helpful discussions. 
The BCI forest dynamics research project was made possible by National Science Foundation grants to Stephen P. Hubbell: DEB–0640386, DEB–0425651, DEB–0346488, DEB–0129874, DEB–00753102, DEB–9909347, DEB–9615226, DEB–9615226, DEB–9405933, DEB–9221033, DEB–9100058, DEB–8906869, DEB–8605042, DEB–8206992, DEB–7922197; support from the Center for Tropical Forest Science; the Smithsonian Tropical Research Institute; the John D.and  Catherine T. MacArthur Foundation; the Mellon Foundation; the Small World Institute Fund; and  numerous private individuals, and through the hard work of over 100 people from 10 countries over the past two decades.
Calculations were carried out using {\tt spatstat} \cite{Baddeley4} and {\tt GET} \cite{Myllymaki}.

\section*{Appendix}

\begin{figure}[hbt]
\centering
\begin{subfigure}{.4\textwidth}
  \centering
  \includegraphics[width=0.8\linewidth]{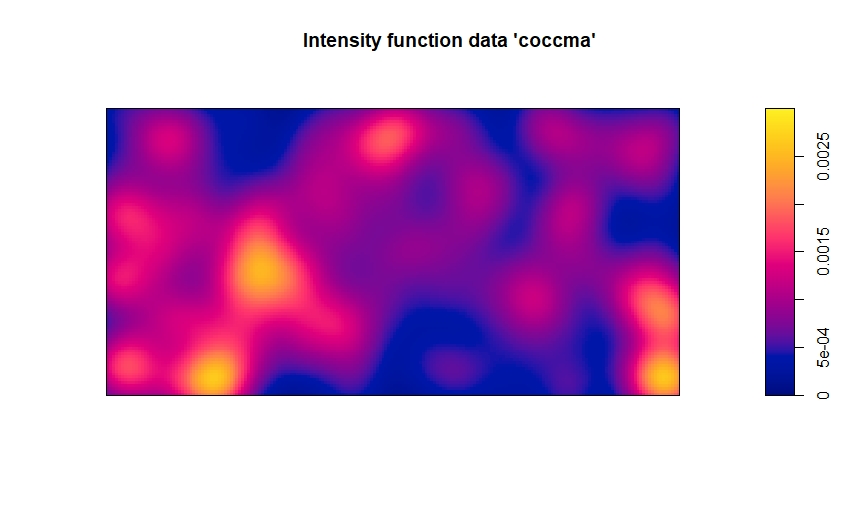}
  \caption{Kernel smoothed intensity function of `coccma'.}
  \label{lambda_coccma}
\end{subfigure}\hspace{0.02\textwidth}
\begin{subfigure}{.4\textwidth}
  \centering
  \includegraphics[width=0.8\linewidth]{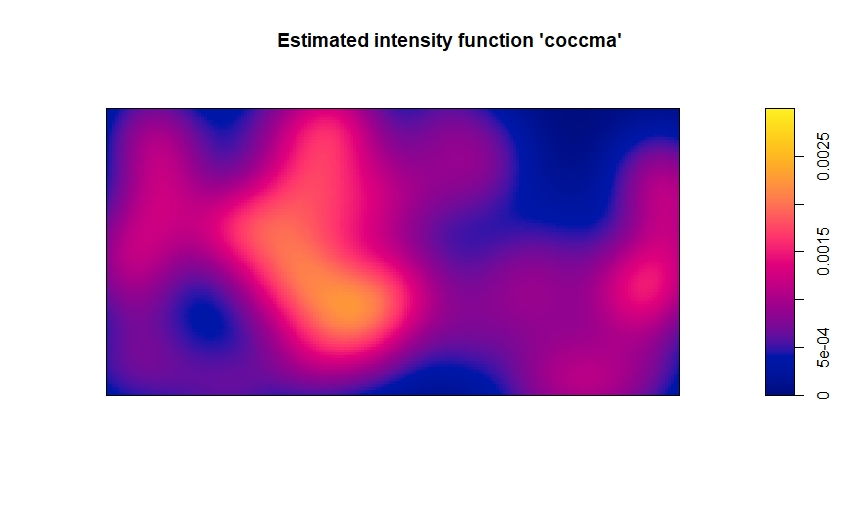}
  \caption{Estimated intensity function for `coccma'.}
  \label{lambda_coccma_sim}
\end{subfigure}
\vspace{2cm}
\begin{subfigure}{.4\textwidth}
  \centering
  \includegraphics[width=0.8\linewidth]{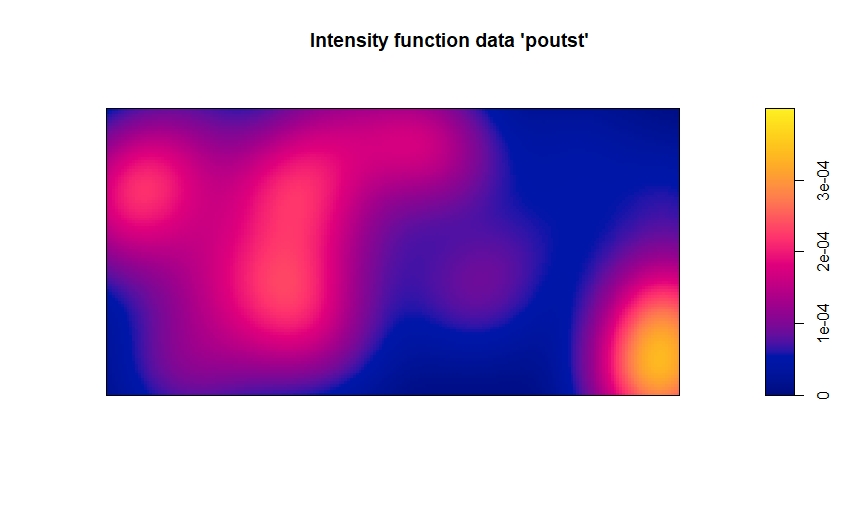}
  \caption{Kernel smoothed intensity function of `poutst'.}
  \label{lambda_poutst}
\end{subfigure}\hspace{0.02\textwidth}
\begin{subfigure}{.4\textwidth}
  \centering
  \includegraphics[width=0.8\linewidth]{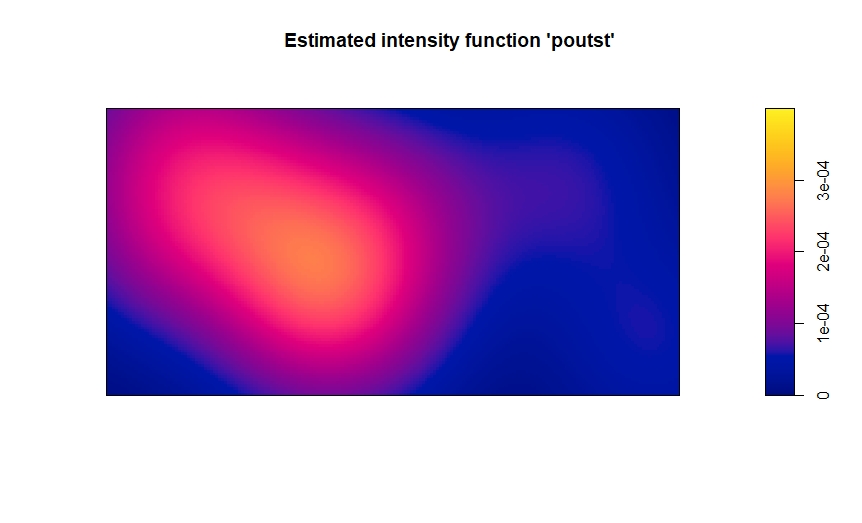}
  \caption{Estimated intensity function for `poutst'.}
  \label{lambda_poutst_sim}
\end{subfigure}\\
\caption{Comparison of estimated intensity functions.}
\end{figure}

\clearpage

\begin{figure}
\centering
\begin{subfigure}{.45\textwidth}
  \centering
  \includegraphics[width=0.8\linewidth]{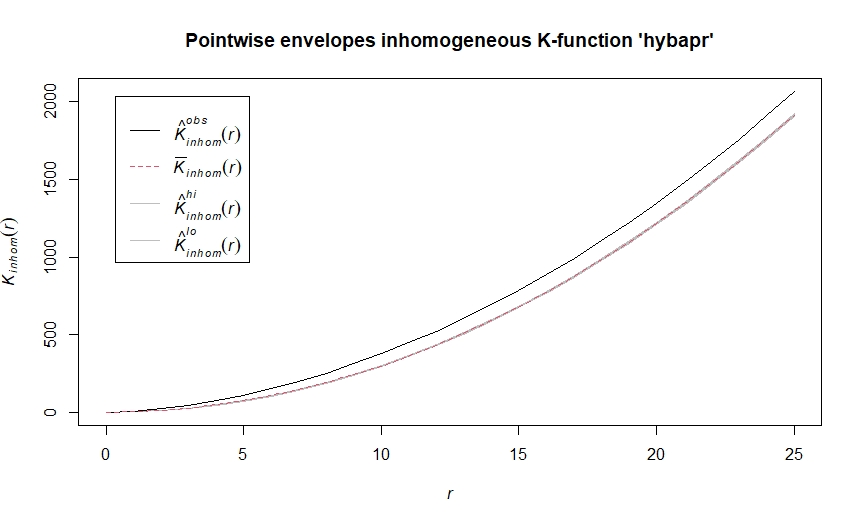}
  \caption{Pointwise envelopes of estimated inhomogeneous $K$-function of `hybapr'.}
  \label{Kenvelopes_hybapr}
\end{subfigure}

\begin{subfigure}{.45\textwidth}
  \centering
  \includegraphics[width=0.8\linewidth]{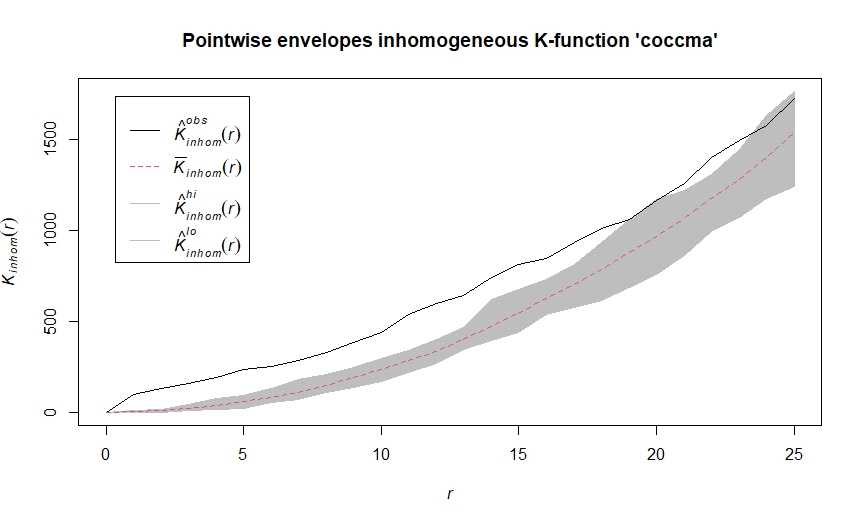}
  \caption{Pointwise envelopes of estimated inhomogeneous $K$-function of `coccma'.}
  \label{Kenvelopes_coccma}
\end{subfigure}

\begin{subfigure}{.45\textwidth}
  \centering
  \includegraphics[width=0.8\linewidth]{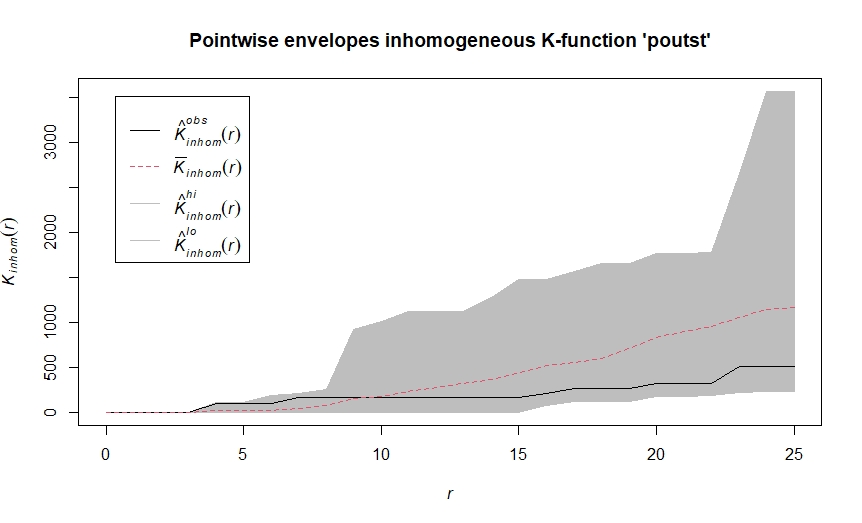}
  \caption{Pointwise envelopes of estimated inhomogeneous $K$-function of `poutst'.}
  \label{Kenvelopes_poutst}
\end{subfigure}
\caption{Simulation envelopes for estimated inhomogeneous $K$-functions based on
$19$ simulations.}
\label{Kenvelopes}
\end{figure}

\clearpage

\begin{figure}
\centering
\begin{subfigure}{.45\textwidth}
  \centering
  \includegraphics[width=0.8\linewidth]{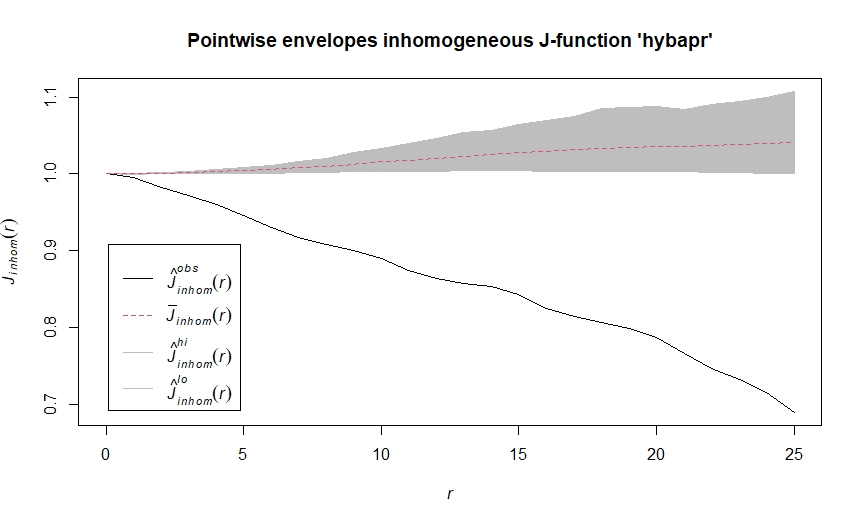}
  \caption{Pointwise envelopes of estimated inhomogeneous $J$-function of `hybapr'.}
  \label{Jenvelopes_hybapr}
\end{subfigure}

\begin{subfigure}{.45\textwidth}
  \centering
  \includegraphics[width=0.8\linewidth]{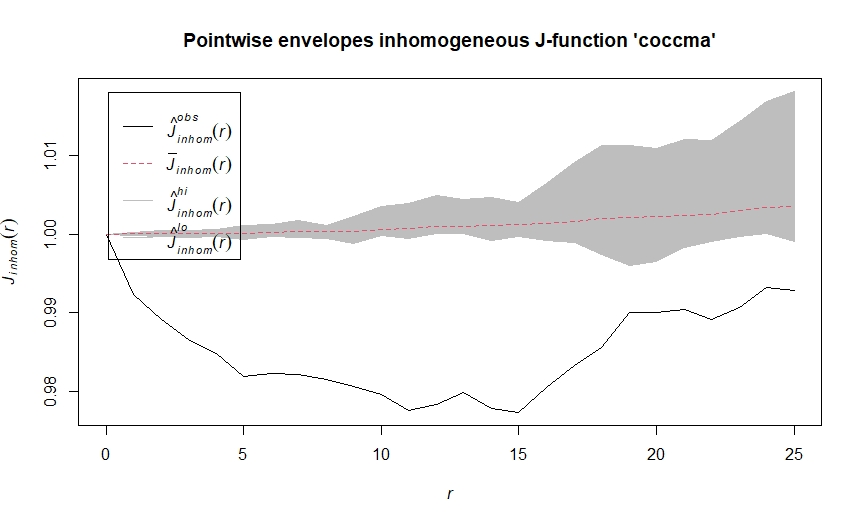}
  \caption{Pointwise envelopes of estimated inhomogeneous $J$-function of `coccma'.}
  \label{Jenvelopes_coccma}
\end{subfigure}\hspace{0.05\textwidth}

\begin{subfigure}{.45\textwidth}
  \centering
  \includegraphics[width=0.8\linewidth]{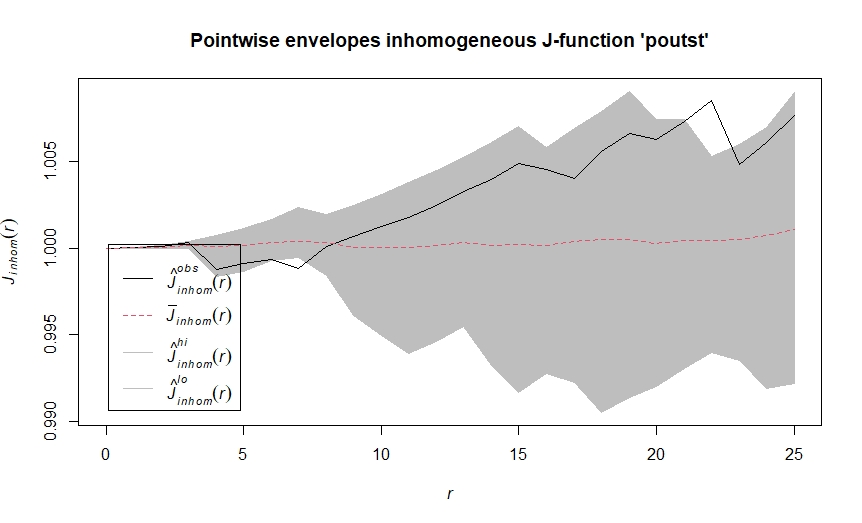}
  \caption{Pointwise envelopes of estimated inhomogeneous $J$-function of `poutst'.}
  \label{Jenvelopes_poutst}
\end{subfigure}\hspace{0.02\textwidth}
\caption{Simulation envelopes for estimated inhomogeneous $J$-functions based on
$19$ simulations.}
\label{Jenvelopes}
\end{figure}

\clearpage

\begin{table}
\begin{center}
 \begin{tabular}{||c||c||c||c||c||} 
 \hline
 Summary & MAD test & DCLF test & Studentized & Directional quantile \\
  statistic  & & & MAD test & MAD test \\
 \hline
 \hline
 K & 0.01 & 0.01 & 0.01 & 0.01\\
 J & 0.01 & 0.01 & 0.01 & 0.01\\
 \hline  
 \end{tabular}    
\caption{$P$-values of the MAD test, the DCLF test, the studentized MAD test and the directional quantile MAD test for `hybapr'.}
\end{center}
\end{table}

\begin{table}
\begin{center}
 \begin{tabular}{||c||c||c||c||c||} 
 \hline
 Summary & MAD test & DCLF test & Studentized & Directional quantile \\
  statistic  & & & MAD test & MAD test \\
 \hline
 \hline
 K & 0.11 & 0.02 & 0.01 & 0.01\\
 J & 0.01 & 0.01 & 0.01 & 0.01\\
 \hline  
 \end{tabular}    
\caption{$P$-values of the MAD test, the DCLF test, the studentized MAD test and the directional quantile MAD test for `coccma'.}
\end{center}
\end{table}

\begin{table}
\begin{center}
 \begin{tabular}{||c||c||c||c||c||} 
 \hline
 Summary & MAD test & DCLF test & Studentized & Directional quantile \\
  statistic  & & & MAD test & MAD test \\
 \hline
 \hline
 K & 0.56 & 0.73 & 0.32 & 0.65\\
 J & 0.49 & 0.96 & 0.14 & 0.15\\
 \hline  
 \end{tabular}    
\caption{$P$-values of the MAD test, the DCLF test, the studentized MAD test and the directional quantile MAD test for `poutst'.}
\end{center}
\end{table}

\clearpage

\begin{figure}
\centering
\begin{subfigure}{.45\textwidth}
  \centering
  \includegraphics[width=0.8\linewidth]{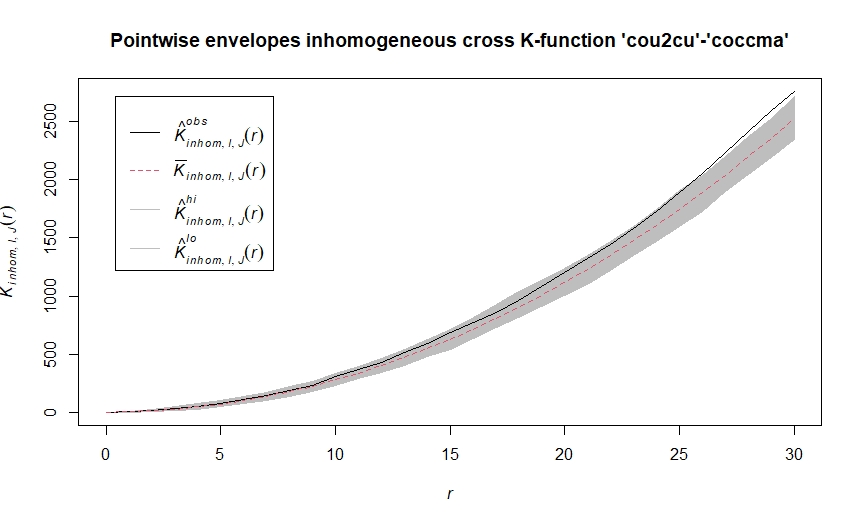}
  \caption{Pointwise envelopes of estimated inhomogeneous cross $K$-function of `cou2cu' -- `coccma'.}
\end{subfigure}\hspace{0.05\textwidth}
\begin{subfigure}{.45\textwidth}
  \centering
  \includegraphics[width=0.8\linewidth]{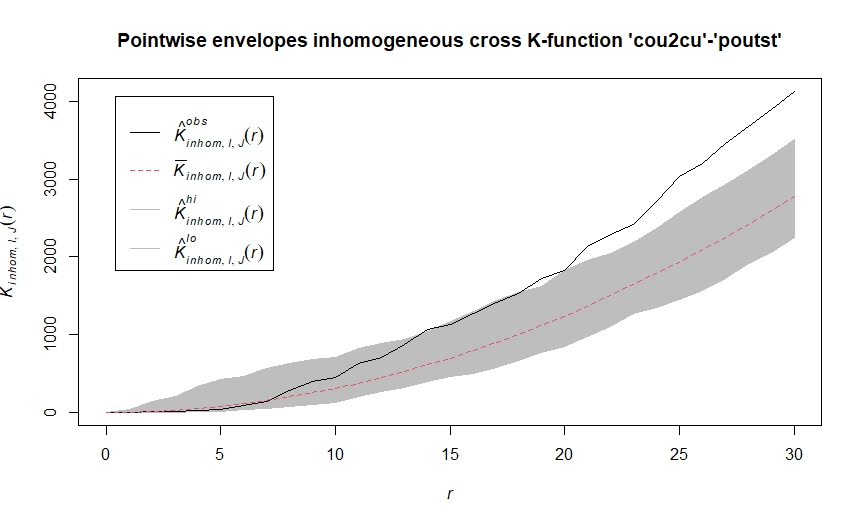}
  \caption{Pointwise envelopes of estimated inhomogeneous cross $K$-function of `cou2cu' -- `poutst'.}
\end{subfigure}\hspace{0.02\textwidth}
\begin{subfigure}{.45\textwidth}
  \centering
  \includegraphics[width=0.8\linewidth]{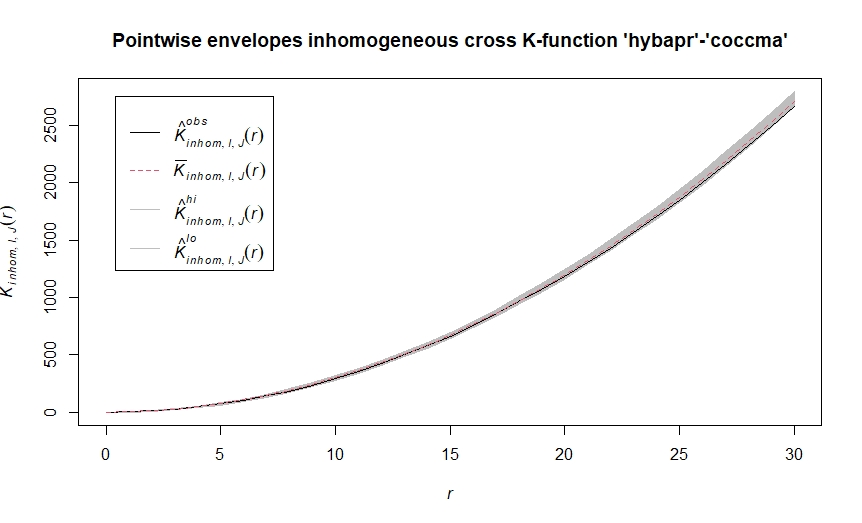}
  \caption{Pointwise envelopes of estimated inhomogeneous cross $K$-function of `hybapr' -- `coccma'.}
\end{subfigure}\hspace{0.05\textwidth}
\begin{subfigure}{.45\textwidth}
  \centering
  \includegraphics[width=0.8\linewidth]{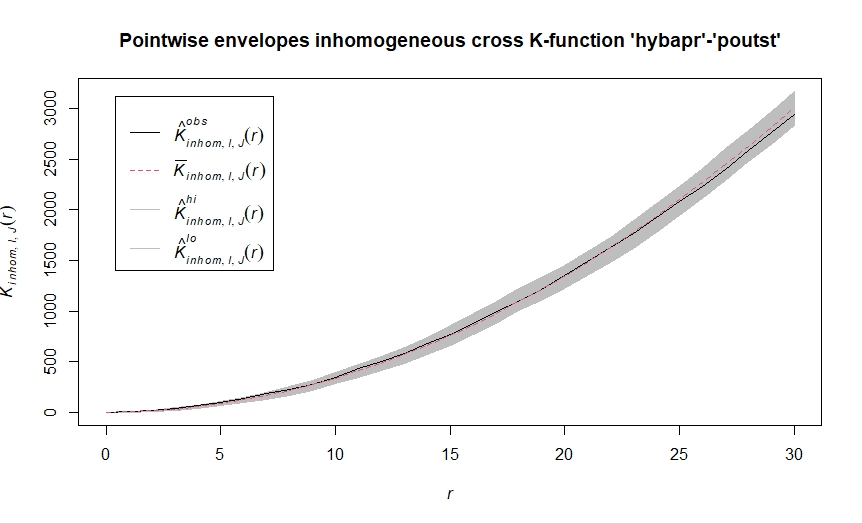}
  \caption{Pointwise envelopes of estimated inhomogeneous cross $K$-function of `hybapr' -- `poutst'.}
\end{subfigure}\hspace{0.02\textwidth}
\begin{subfigure}{.45\textwidth}
  \centering
  \includegraphics[width=0.8\linewidth]{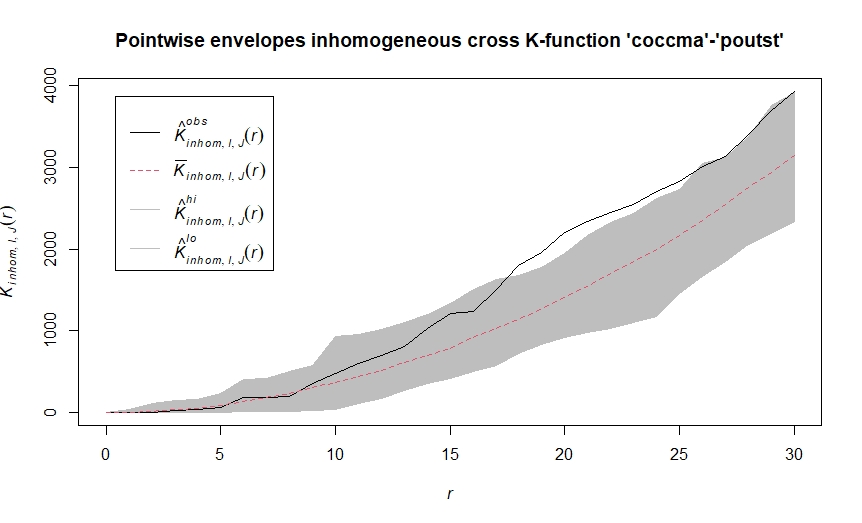}
  \caption{Pointwise envelopes of estimated inhomogeneous cross $K$-function of `coccma' -- `poutst'.}
\end{subfigure}\hspace{0.02\textwidth}
\caption{Simulation envelopes for estimated inhomogeneous cross $K$-functions based on $99$ simulations.}
\end{figure}

\clearpage

\begin{figure}
\centering
\begin{subfigure}{.45\textwidth}
  \centering
  \includegraphics[width=0.8\linewidth]{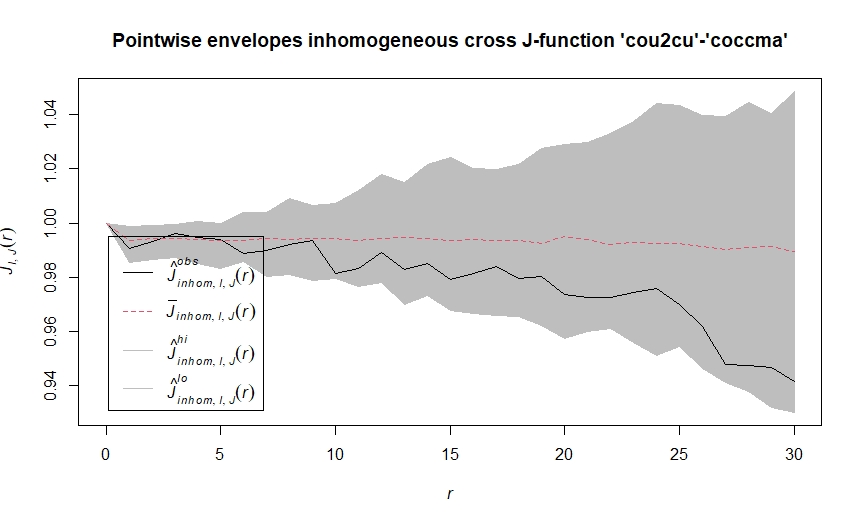}
  \caption{Pointwise envelopes of estimated inhomogeneous cross $J$-function of `cou2cu' -- `coccma'.}
\end{subfigure}\hspace{0.05\textwidth}
\begin{subfigure}{.45\textwidth}
  \centering
  \includegraphics[width=0.8\linewidth]{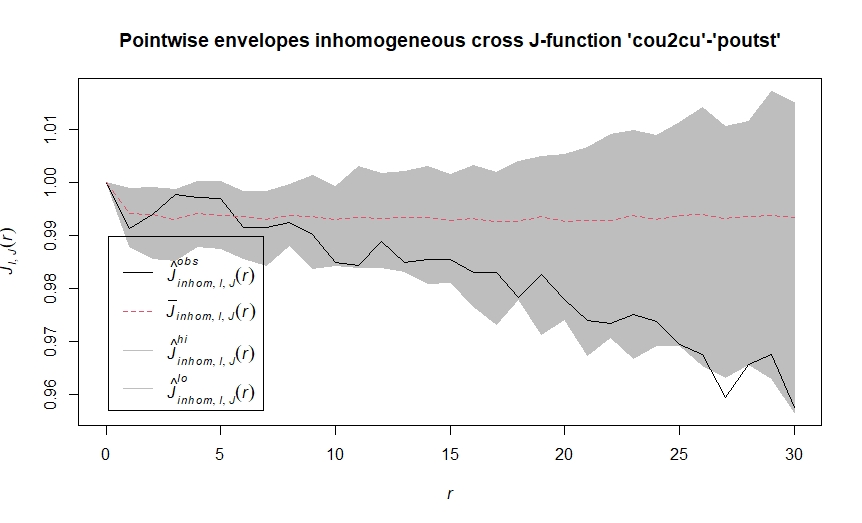}
  \caption{Pointwise envelopes of estimated inhomogeneous cross $J$-function of `cou2cu' -- `poutst'.}
\end{subfigure}\hspace{0.02\textwidth}
\begin{subfigure}{.45\textwidth}
  \centering
  \includegraphics[width=0.8\linewidth]{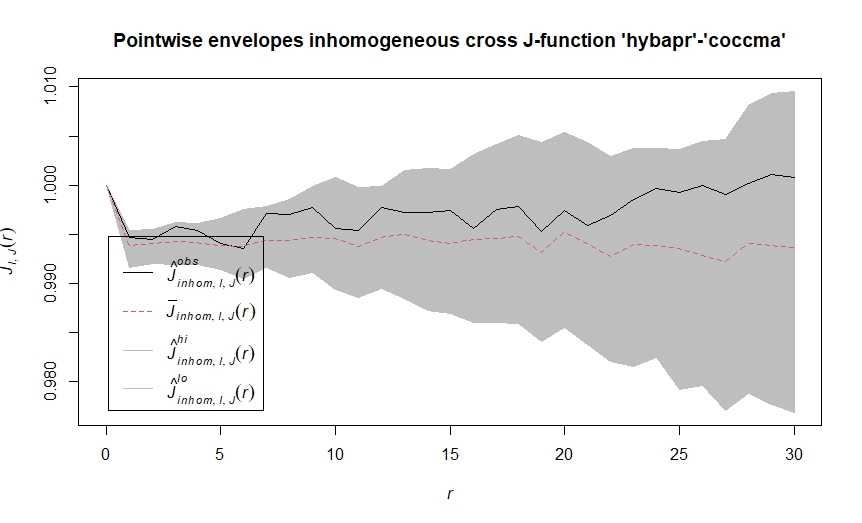}
  \caption{Pointwise envelopes of estimated inhomogeneous cross $J$-function of `hybapr'-- `coccma'.}
\end{subfigure}\hspace{0.05\textwidth}
\begin{subfigure}{.45\textwidth}
  \centering
  \includegraphics[width=0.8\linewidth]{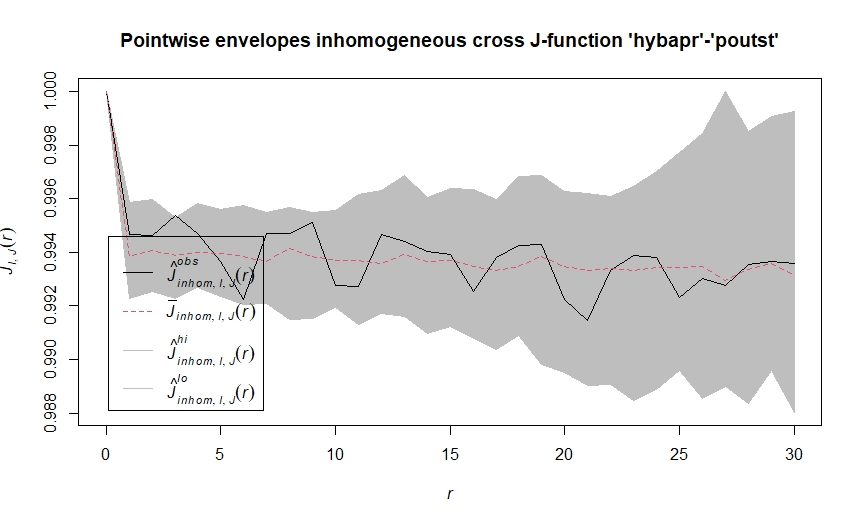}
  \caption{Pointwise envelopes of estimated inhomogeneous cross $J$-function of `hybapr'-- `poutst'.}
\end{subfigure}\hspace{0.02\textwidth}
\begin{subfigure}{.45\textwidth}
  \centering
  \includegraphics[width=0.8\linewidth]{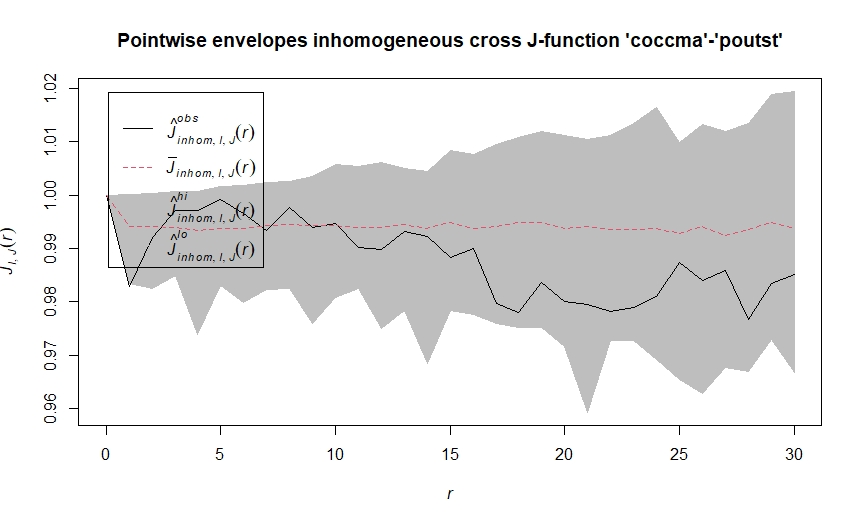}
  \caption{Pointwise envelopes of estimated inhomogeneous cross $J$-function of `coccma' -- `poutst'.}
\end{subfigure}\hspace{0.02\textwidth}
\caption{Simulation envelopes for estimated inhomogeneous cross $J$-functions based on $99$ simulations.}
\end{figure}

\clearpage

\begin{table}
\begin{center}
 \begin{tabular}{||c||c||c||c||c||} 
 \hline
 Summary statistic & MAD test & DCLF test & Studentized & Directional quantile \\
  & & & MAD test & MAD test \\
 \hline
 \hline
 K & 0.01 & 0.02 & 0.02 & 0.03\\
 J & 0.06 & 0.09 & 0.29 & 0.31\\
 \hline  
 \end{tabular}    
\caption{$P$-values of the MAD test, the DCLF test, the studentized MAD test and the directional quantile MAD test for `cou2cu' -- `coccma'.}
\end{center}
\end{table} 

\begin{table}
\begin{center}
 \begin{tabular}{||c||c||c||c||c||} 
 \hline
 Summary statistic & MAD test & DCLF test & Studentized & Directional quantile \\
  & & & MAD test & MAD test \\
 \hline
 \hline
 K & 0.01 & 0.01 & 0.02 & 0.02 \\
 J & 0.02 & 0.02 & 0.02 & 0.01\\
 \hline  
 \end{tabular}    
\caption{$P$-values of the MAD test, the DCLF test, the studentized MAD test and the directional quantile MAD test for `cou2cu' -- `poutst'.}
\end{center}
\end{table} 

\begin{table}
\begin{center}
 \begin{tabular}{||c||c||c||c||c||} 
 \hline
 Summary statistic & MAD test & DCLF test & Studentized & Directional quantile \\
  & & & MAD test & MAD test \\
 \hline
 \hline
 K & 0.08 & 0.16 & 0.32 & 0.23 \\
 J & 0.50 & 0.38 & 0.42 & 0.35\\
 \hline  
 \end{tabular}    
\caption{$P$-values of the MAD test, the DCLF test, the studentized MAD test and the directional quantile MAD test for `hybapr'-- `coccma'.}
\end{center}
\end{table} 

\begin{table}
\begin{center}
 \begin{tabular}{||c||c||c||c||c||} 
 \hline
 Summary statistic & MAD test & DCLF test & Studentized & Directional quantile \\
  & & & MAD test & MAD test \\
 \hline
 \hline
 K & 0.47 & 0.74 & 0.59 & 0.61 \\
 J & 0.83 & 0.78 & 0.21 & 0.29\\
 \hline  
 \end{tabular}    
\caption{$P$-values of the MAD test, the DCLF test, the studentized MAD test and the directional quantile MAD test for `hybapr'-- `poutst'.}
\end{center}
\end{table} 

\begin{table}
\begin{center}
 \begin{tabular}{||c||c||c||c||c||} 
 \hline
 Summary statistic & MAD test & DCLF test & Studentized & Directional quantile \\
  & & & MAD test & MAD test \\
 \hline
 \hline
 K & 0.04 & 0.01 & 0.08 & 0.06\\
 J & 0.40 & 0.11 & 0.12 & 0.06\\
 \hline  
 \end{tabular}    
\caption{$P$-values of the MAD test, the DCLF test, the studentized MAD test and the directional quantile MAD test for `coccma' -- `poutst'.}
\end{center}
\end{table} 

\clearpage

\bibliography{barro.bib}

\begin{thebibliography}{10}

\bibitem{Baddeley4}
A.~Baddeley, E.~Rubak, and R.~Turner.
\newblock {\em Spatial point patterns. Methodology and applications with R}.
\newblock CRC Press, 2016.

\bibitem{Baddeley1}
A.J. Baddeley, J.~M\o ller, and R.~Waagepetersen.
\newblock Non- and semi-parametric estimation of interaction in inhomogeneous
  point patterns.
\newblock {\em Statistica Neerlandica}, 54:329--350, 2000.

\bibitem{Chesson}
P.~Chesson.
\newblock General theory of competitive coexistence in spatially-varying
  environments.
\newblock {\em Theoretical Population Biology}, 58:211--237, 2000.

\bibitem{Clements}
F.E. Clements.
\newblock {\em Plant succession: an analysis of the development of vegetation}.
\newblock Carnegie Institute of Washington, 1916.

\bibitem{Cronie1}
O.~Cronie and M.N.M. van Lieshout.
\newblock Summary statistics for inhomogeneous marked point processes.
\newblock {\em Annals of the Institute of Statistical Mathematics},
  68:905--928, 2015.

\bibitem{Cronie2}
O.~Cronie and M.N.M. van Lieshout.
\newblock A non-model-based approach to bandwidth selection for kernel
  estimators of spatial intensity functions.
\newblock {\em Biometrika}, 105:455--462, 2018.

\bibitem{Diggle2}
P.~Diggle.
\newblock A kernel method for smoothing point process data.
\newblock {\em Journal of the Royal Statistical Society Series C (Applied
  Statistics)}, 34:138--147, 1985.

\bibitem{Diggle1}
P.J. Diggle.
\newblock {\em Statistical analysis of spatial point patterns}.
\newblock Academic Press, 1983.

\bibitem{Gause}
G.F. Gause.
\newblock {\em The struggle for existence}.
\newblock Williams \& Wilkins, 1934.

\bibitem{Gentry}
A.H. Gentry.
\newblock Tree species richness of upper {Amazonian} forests.
\newblock {\em Proceedings of the National Academy of Sciences of the United
  States of America}, 85:156--159, 1988.

\bibitem{Gleason}
H.A. Gleason.
\newblock The individualistic concept of the plant association.
\newblock {\em Bulletin of the Torrey Botanical Club}, 53:7--26, 1926.

\bibitem{Hubbell1}
S.P. Hubbell.
\newblock {\em The unified neutral theory of biodiversity and biogeography}.
\newblock Princeton University Press, 2001.

\bibitem{Hubbell2}
S.P. Hubbell and R.B. Foster.
\newblock {\em Diversity of canopy trees in a neotropical forest and
  implications for conservation}, pages 25--41.
\newblock Blackwell, 1983.

\bibitem{Huisman}
J.~Huisman and F.J. Weissing.
\newblock Biodiversity of plankton by species oscillations and chaos.
\newblock {\em Nature}, 402:407--410, 1999.

\bibitem{Lieberman}
M.~Lieberman and D.~Lieberman.
\newblock Nearest-neighbor tree species combinations in tropical forests: the
  role of chance, and some consequences of high diversity.
\newblock {\em Oikos}, 116:377--386, 2007.

\bibitem{VanLieshout2}
{M.N.M.~van} Lieshout.
\newblock A {J-function} for inhomogeneous point processes.
\newblock {\em Statistica Neerlandica}, 65:183--201, 2011.

\bibitem{VanLieshout3}
{M.N.M.~van} Lieshout.
\newblock On estimation of the intensity function of a point process.
\newblock {\em Methodology and Computing in Applied Probability}, 14:567--578,
  2012.

\bibitem{VanLieshout1}
{M.N.M.~van} Lieshout.
\newblock Nonparametric indices of dependence between components for
  inhomogeneous multivariate random measures and marked sets.
\newblock {\em Scandinavian Journal of Statistics}, 45:985--1015, 2018.

\bibitem{MollerWaagepetersen}
J.~M\o ller and R.P. Waagepetersen.
\newblock {\em Statistical inference and simulation for spatial point
  processes}.
\newblock CRC Press, 2004.

\bibitem{Losos}
E.C. Losos and E.G. Leigh, editors.
\newblock {\em Tropical forest diversity and dynamism: findings from a
  large-scale plot network}.
\newblock University of Chicago Press, 2004.

\bibitem{Lotwick}
H.W. Lotwick and B.W. Silverman.
\newblock Methods for analysing spatial processes of several types of points.
\newblock {\em Journal of the Royal Statistical Society, Series B (Statistical
  Methodology)}, 44:406--413, 1982.

\bibitem{McGill}
B.J. McGill.
\newblock Towards a unification of unified theories of biodiversity.
\newblock {\em Ecology Letters}, 13:627--642, 2010.

\bibitem{Myllymaki}
M.~Myllymäki, T.~Mrkvička, P.~Grabarnik, H.~Seijo, and U.~Hahn.
\newblock Global envelope tests for spatial processes.
\newblock {\em Journal of the Royal Statistical Society Series B (Statistical
  Methodology)}, 79:381--404, 2016.

\bibitem{Perry}
G.L.W. Perry, N.J. Enright, B.P. Miller, and B.P. Lamont.
\newblock Nearest-neighbour interactions in species-rich shrublands: the roles
  of abundance, spatial patterns and resources.
\newblock {\em Oikos}, 118:161--174, 2009.

\bibitem{Ricklefs}
R.E. Ricklefs.
\newblock {\em Ecology}.
\newblock W.H. Freeman and Company, 3rd edn. edition, 1990.

\bibitem{Tilman}
D.~Tilman.
\newblock {\em Resource competition and community structure}.
\newblock Princeton University Press, 1982.

\bibitem{Volkov}
I.~Volkov, J.R. Banavar, S.P. Hubbell, and A.~Maritan.
\newblock Inferring species interactions in tropical forests.
\newblock {\em Proceedings of the National Academy of Sciences of the United
  States of America}, 106:13854--13859, 2009.

\bibitem{Waagepetersen}
R.~Waagepetersen, Y.~Guan, A.~Jalilian, and J.~Mateu.
\newblock Analysis of multispecies point patterns by using multivariate
  log-{Gaussian} {Cox} processes.
\newblock {\em Journal of the Royal Statistical Society Series C (Applied
  Statistics)}, 65:77--96, 2016.

\bibitem{Wiegand2}
T.~Wiegand, C.V.S. Gunatilleke, I.A.U.N. Gunatilleke, and A.~Huth.
\newblock How individual species structure diversity in tropical forests.
\newblock {\em Proceedings of the National Academy of Sciences of the United
  States of America}, 104:19029--19033, 2007.

\bibitem{Wiegand3}
T.~Wiegand, A.~Huth, S.~Getzin, X.~Wang, Z.~Hao, C.V.S. Gunatilleke, and
  I.A.U.N. Gunatilleke.
\newblock Testing the independent species' arrangement assertion made by
  theories of stochastic geometry of biodiversity.
\newblock {\em Proceedings of the Royal Society B}, 279:3312--3320, 2012.

\bibitem{Wright}
S.J. Wright.
\newblock Plant diversity in tropical forests: a review of mechanisms of
  species coexistence.
\newblock {\em Oecologia}, 130:1--14, 2002.

\end{thebibliography}

\end{document}